\documentclass[aps,prl,reprint,groupedaddress]{revtex4-1}
\usepackage{amsmath,amssymb}
\usepackage{graphicx}% Include figure files
\usepackage{dcolumn}% Align table columns on decimal point
\usepackage{bm}% bold math
\usepackage[utf8]{inputenc}
\usepackage[T1]{fontenc}
\usepackage{mathptmx}

\begin{document}

\preprint{PSBJ/123}

\title[]{Pati-Salam model in curved space-time from square root Lorentz manifold}
% Force line breaks with \\

\author{De-Sheng Li}
 \altaffiliation[At ]{College of Physics, Mechanical and Electrical Engineering, Jishou University, Jishou 416000, P.~R.~China}
\altaffiliation[Also at ]{Interdisciplinary Center for Quantum Information, National University of Defense Technology, Changsha 410073, P.~R.~China }
\altaffiliation[Also at ]{Institute of High Energy Physics,
Chinese Academy of Sciences, 19B Yuquan Road,
Beijing 100049, P.~R.~China }
\email[\\E-mail:]{lideshengjy@126.com}

\date{\today}% It is always \today, today,

\begin{abstract}
There is a $U (4^{\prime}) ×U (4)$-bundle on four-dimensional square root Lorentz manifold. Then a Pati-Salam model in curved space-time (Lagrangian) and a gravity theory (Lagrangian) are constructed on square root Lorentz manifold based on self-parallel transportation principle. An explicit formulation of Sheaf quantization on this square root Lorentz manifold is shown. Sheaf quantization is based on superposition principle and construct a linear Sheaf space in curved space-time. The transition amplitude in path integral quantization is given which is consistent with Sheaf quantization. All particles and fields in Standard Model (SM) of particle physics and Einstein gravity are found in square root metric and the connections of bundle. The interactions between particles/fields are described by Lagrangian explicitly. There are few new physics in this model. The gravity theory is Einstein-Cartan kind with torsion. There are new particles, right handed neutrinos, dark photon, Fiona, $X^{±C}$ and $Y^0,Y^1,Y^2,Y_*^1,Y_*^2$.
\end{abstract}

\maketitle

%\begin{widetext}
\section{Introduction}
Four-dimensional pseudo Riemann geometry with signature $(-,+,+,+)$, Lorentz manifold, is the geometry background of the general relativity, space-time is described by the metric, and the gravitational field is described as the curve of space-time. In general relativity, the geodesic equation describes the trajectories of freee particles, and the Einstein equation determines how matter curves space-time. At the last life time of Einstein, he attempted to establish a new geometry unifying electromagnetic interaction and gravity. This idea was developed by Weil into the early idea of gauge invariance and by the Kaluza and Klein into the idea of extra dimensions.

Later, the Yang-Mills theory \cite{Yang1954} was confirmed. Yang-Mills theory takes gauge invariance as its basic principle and to be the theoretical framework of electromagnetic, weak and strong interaction in Standard Model (SM) of particle physics \cite{Glashow:1961tr,Weinberg:1967tq,Salam:1968rm,Gross:1973id,Politzer:1973fx,Langacker:1980js}. The Yang-Mills theory is the theoretical framework of SM and has a good correspondence with the complex structure group G fiber bundle theory \cite{Hou:1999qc}. General relativity can actually be rewritten in the framework of fiber bundle theory also, except that the G structure group of general relativity is real, and the corresponding fiber bundles are the tangent and cotangent bundles.
Another way to build unified field theory is introducing extra dimensions to give all fields their geometric positions. And lots of attempts in extra dimension were made. 
Is it possible to fuse the tangent (cotangent) bundle of general relativity with the complex structure group G-bundle of Yang-Mills theory? 

Inspired by the Dirac's way of finding his equation and spinors through making square root of the Klein-Gordon equation, we researched the square root of the the four-dimensional Lorentz manifold, which similar with the papers in Clifford algebra or Clifford bundle \cite{Dowker:1978vy,Casalbuoni:1979ta,Rodrigues:1993cz,Smith:1994ek,White:1997cc,Keller:1998sdn,Vargas:2001nz,Pavsic:2002fw,Krolikowski:2002jr,Mosna:2002fr,Castro:2002kt,ElNaschie:2004dp,Pavsic:2004ni,Pavsic:2004ima,Pavsic:2005uj,Vacaru:2005ht,Coquereaux:2005up,Castro:2005qz,GarrettLisi:2005gy,Pavsic:2005qs,Castro:2009zzd,Smarandache:2008zzb,Castro:2010zza,Pavsic:2010zz,Castro:2012zza,Castro:2013sc,CarlosCastro:2013zgs,Rodrigues:2013ila,Castro:2014wna,Castro:2015jky,HWANG:2016cln,Castro:2016rvj,Stoica:2017iuo,Stoica:2017mim,Pavsic:2017ebn,Lu:2017kvy,Kurkov:2017wmx,Gu:2018bww,Gresnigt:2019orj,Gresnigt:2020jvo,Lukman:2020fhx,Gresnigt:2020gbp,Trindade:2020wrv,Sogami:2020lad}, spin-gauge theory in Riemann-Cartan space-time \cite{Drechsler:1987tf,Dehnen:1994wb}, sedenion \cite{Weng:2018riu} and Einstein-Cartan theory \cite{hehl1976general,Tecchiolli:2019hfe, Cebecioglu:2021dqb} etc. Four-dimensional square root Lorentz manifold has extra $U(4^{\prime})\times U(4)$ principal bundle than Lorentz manifold. Two Lagrangians based on four-dimension square root Lorentz manifold are constructed which describe a $U(4^{\prime})\times U(4)_{L}\times U(4)_{R}$ Pati-Salam model in curved space-time and a gravity theory, respectively. In the Pati-Salam model \cite{Pati:1974yy}, the $SU(4^{\prime})$ is color group with ``lepton number as the fourth color'', and the $SU(4)_{L}\times SU(4)_{R}$ is chiral flavor group.
We realize an explicit formulation of Sheaf quantization \cite{Baez:1999sr,Isham:1999kb,Froyshov:2002qaq,Zois:2005ut,Doring:2007ib,Doring:2007ic,Doring:2007id,Fukaya:2011ad,Bochicchio:2013aha,Cho:2013xda,Zois:2014wka,Nakayama:2014tga,2016LMaPh106,2018CMaPh,Viterbo2019,Kuwagaki2020pry,Asano2020} scheme which consistents with path integral quantization. The particles spectrum on this model is discussed.

%%
%% TABLES
%%
%% If there are any tables, put them here.
%%

%% Put the bibliography here, most people will use BiBTeX in
%% which case the environment below should be replaced with
%% the \bibliography{} command.

%\begin{thebibliography}{1}
%\bibitem{dummy} Articles are restricted to 50 references, Letters
%to 30.
%\bibitem{dummyb} No compound references -- only one source per
%reference.
%\end{thebibliography}

%\appendix

\section{Geometry and Lagrangian}
\label{sec:1}

The notations are introduced here. $a,b,c,d$ represent frame indices, and $a,b,c,d=0,1,2,3$. $\mu ,\nu, \rho, \sigma$ represent coordinates indices, and $\mu ,\nu, \rho, \sigma=0,1,2,3$. $\alpha$ represent group indices, and $\alpha=0,1 ,\cdots,15$. $i,j,k,l,m=1,2,3,4$. $C=R,G,B=1,2,3$ is quarks color. $\kappa$ is Sheaf space index. Repeated indices are summed by default. 

%and \cite{RefJ}
The pseudo Riemann manifold is described by a metric
\begin{eqnarray}
g(x) &=& -g_{\mu \nu}(x){{dx^{\mu}}\otimes{dx^{\nu}}},\\
 g_{\mu \nu}(x)= g_{\nu \mu}(x),&\quad&g_{v}=\det (g_{\mu \nu}(x)) ,
\end{eqnarray}
where $\{x|x=(x^{\mu})=(t,\vec{x})\}$ is a four-dimensional topological space. Here we discuss the four-dimensional pseudo Riemann manifold with signature $(-,+,+,+)$, Lorentz manifold. 
And it can be described by orthonormal frame formalism as
\begin{equation}\label{g}
g^{-1}(x) =-{\eta }^{ab}{{\theta }_{a}}(x){{\theta }_{b}}(x), 
\end{equation}
where $\eta^{ab}=diag(1,-1,-1,-1)$ and ${\theta }_{a}(x)={\theta }_{a}^{\mu}(x)\frac{\partial}{\partial x^{\mu}}$ are orthonormal frames and describe gravitational field.

The definition of gamma matrices is
\begin{equation}\label{clifford}
    {{\gamma }^{a}}{{\gamma }^{b}}+{{\gamma }^{b}}{{\gamma }^{a}}=2\eta^{ab}I_{4\times 4}.
 \end{equation}
The Hermiticity conditions for gamma matrices are
\begin{equation}\label{cliffordA}
    {{\gamma }^{a}}{{\gamma }^{b\dagger}}+{{\gamma }^{b\dagger}}{{\gamma }^{a}}=2I^{ab}I_{4\times 4},
 \end{equation}
 where $I^{ab}=diag(1,1,1,1)$.
 We define
\begin{eqnarray}\label{lx}
l(x)&=&i\gamma^{0}_{ik}(x){\gamma }^{a}_{kj}(x)e^{\dagger}_{j}\otimes e_{i}{{\theta }_{a}}(x) ,\\
\tilde{l}(x)&=&i{\gamma }^{a}_{ik}(x)\gamma^{0}_{kj}(x)e^{\dagger}_{j}\otimes e_{i}{{\theta }_{a}}(x),
\end{eqnarray}
where $e_{i}$ are the orthogonal bases expanding four-dimension complex space $\mathbb{C}^{4}$ and 
\begin{equation}\label{ei}
\mathbf{tr}( e^{\dagger}_{j} \otimes e_{i})=e_{i}e^{\dagger}_{j}=\delta_{ij}.
\end{equation}
 One simple choice of $e_{i}$ is
\begin{eqnarray} \label{bases} 
e_{1}=(e^{i\theta_{1}},0,0,0), &\quad& e_{2}=(0,e^{i\theta_{2}},0,0),\\
e_{3}=(0,0,e^{i\theta_{3}},0), &\quad& e_{4}=(0,0,0,e^{i\theta_{4}}).
\end{eqnarray}

After using $\gamma^{a\dagger}=\gamma^{0}\gamma^{a}\gamma^{0}$, we find that
\begin{equation}\label{gl}
 g^{-1}(x)=\frac{1}{4}tr[\tilde{l}(x)l(x)] .
\end{equation}
Then $l(x)$ or $\tilde{l}(x)$ are the square root of metric (\ref{g}) in some sense. The $\gamma^{a}_{ij}(x)$ can be showed as
\begin{eqnarray}\nonumber
  \gamma^{0\prime}_{ik}\gamma^{a\prime}_{kj}(x)&=&\psi^{\dagger}_{ik}(x) \gamma^{0}_{kl}\gamma^{a}_{lm} \psi_{mj}(x)=\bar{\psi}_{i}(x) \gamma^{a} \psi_{j}(x),\\  \nonumber
  \gamma^{a\prime}_{ik}\gamma^{0\prime}_{kj}&=&\psi^{\dagger}_{ik}(x) \gamma^{a}_{kl}\gamma^{0}_{lm} \psi_{mj}(x)=\bar{\psi}_{i}(x) \gamma^{a\dagger} \psi_{j}(x),
\end{eqnarray}
where $\psi_{i}$ are the Dirac fermions field with flavor related index $i=1,2,3,4$.  $\bar{\psi}_{i}(x)=\psi^{\dagger}_{i}(x)\gamma^{0}$, $\psi(x)\in U(4)$ is $4\times 4$ matrix. So, the square root metric are defined as follow
\begin{eqnarray}\label{l}
l(x)=i\bar{\psi}_{i}(x) \gamma^{a} \psi_{j}(x) e^{\dagger}_{j} \otimes e_{i}{{\theta }_{a}}(x),\\   \label{lt}
\tilde{l}(x)=i\bar{\psi}_{i}(x) \gamma^{a\dagger} \psi_{j}(x) e^{\dagger}_{j} \otimes e_{i}{{\theta }_{a}}(x).
\end{eqnarray}
The square root Lorentz manifold is described by square root metric (\ref{l}), (\ref{lt}). 
Direct calculations show that the definition (\ref{l}) and (\ref{lt}) satisfy  (\ref{gl}) and $l^{\dagger}(x)=-l(x), \tilde{l}^{\dagger}(x)=-\tilde{l}(x)$. 
\begin{figure}
\begin{center}
  \includegraphics[width=72 mm]{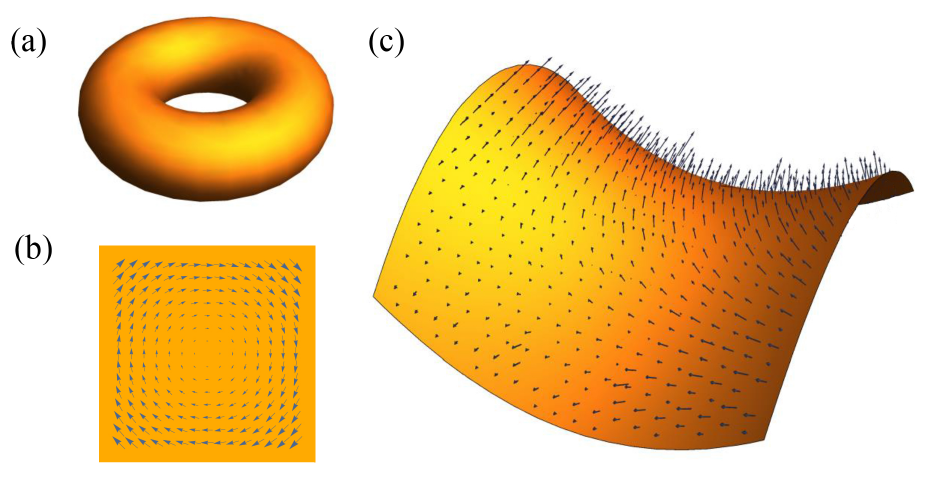}
  \caption{\label{figure1}Picture visualization of geometry background of (a) general relativity, (b) Yang-Mills theory and (c) square root metric. (a) Geometry background of general relativity is pseudo Riemann geometry, which is a smooth curved manifold. (b) Geometry background of Yang-Mills theory is complex G-bundle with flat base manifold, G is the gauge group of Yang-Mills theory. (c) Square root metric geometry in four-dimension has everything of pseudo Riemann geometry and extra $U(4^{\prime})\times U(4)$-bundle. }
\end{center}
\end{figure}

The coefficients of the affine connections on coordinates, coefficients of spin connections on orthonormal frame \cite{Chern:1999jn} and gauge fields on the $U(4^{\prime})\times U(4)$-bundle are defined as follows
%\begin{subequations}
\label{connection}
\begin{equation}\label{connectionA}
\nabla_{\mu} \partial_{\nu}={\Gamma}_{\ \nu\mu}^{\rho}(x)\partial_{\rho},\\
\end{equation}
\begin{equation}\label{connectionB}
    \nabla_{\mu}{{\theta }_{a}}(x)={\Gamma}_{\ a\mu}^{b}(x){{\theta }_{b}}(x),\\
    \end{equation}
    \begin{equation}\label{connectionD}
  \nabla_{\mu} (\gamma^{0}\gamma^{a})=i [V_{\mu }(x)\gamma^{0}\gamma^{a}-\gamma^{0}\gamma^{a}V_{\mu}(x)],
\end{equation}
    \begin{equation}\label{connectionC}
  \nabla_{\mu}e^{\dagger}_{i}=i {W}_{\mu ij}(x)e^{\dagger}_{j},
\end{equation}
%\end{subequations}
where $\Gamma^{b}_{\ a\mu}(x)\theta_{b}^{\rho}(x)=\partial_{\mu}\theta_{a}^{\rho}(x)+\theta_{a}^{\nu}(x)\Gamma^{\rho}_{\ \nu\mu}(x)$ is found and $V_{\mu}^{\dagger}(x)=V_{\mu}(x), W^{*}_{\mu ij}(x)=W_{\mu ji}(x)$. The uniqueness of definition of gauge fields is originated from restriction (\ref{clifford}), (\ref{cliffordA}) and (\ref{ei}). 
The equation as follow can be derived from (\ref{connectionD})
    \begin{equation}\label{connectionE}
  \nabla_{\mu} (\gamma^{a}\gamma^{0})=i [\tilde{V}_{\mu }(x)\gamma^{a}\gamma^{0}-\gamma^{a}\gamma^{0}\tilde{V}_{\mu}(x)],
\end{equation}
where $\tilde{V}_{\mu}(x)=\gamma^{0}V_{\mu}(x)\gamma^{0}$.
The gauge field $V_{\mu}(x)$ and $W_{\mu ij}(x)$ can be decomposed by the generators of the $U(4)$ group   
\begin{eqnarray}\label{gaugeB}
V_{\mu}(x)&=&V^{\alpha}_{\mu}(x) \mathcal{T}^{\alpha},\\  \label{gaugeA}
  W_{\mu ij}(x)&=&W^{\alpha}_{\mu}(x)\mathcal{T}^{\alpha}_{ij},
\end{eqnarray}
where $\alpha=0,1,2,\cdots ,15$ and $V^{\alpha \dagger}_{\mu}(x)=V^{\alpha}_{\mu}(x), W^{\alpha\dagger}_{\mu}(x)=W^{\alpha}_{\mu }(x)$ are gauge bosons fields.
 The $\mathcal{T}^{\alpha}$ are the generators of $U(4)$ and an expilict one can be seen in appendix. A equation which satisfying the $U(4^{\prime})\times U(4)$ gauge invariant, locally Lorentz invariant and generally covariant principles is constructed
\begin{equation}\label{fundamentalA0}
\mathbf{tr}\nabla [l(x)]= 0,
\end{equation}
this equation originated from generalized self parallel transportation principle.
Eliminating index $x$, the explicit formula of equation (\ref{fundamentalA0}) is
%\begin{widetext}
\begin{eqnarray*} %\label{fundamentalA4}  
\left[(i\partial_{\mu}\bar{\psi}_{i}-\bar{\psi}_{i} \tilde{V}_{\mu}+W_{\mu ij}\bar{\psi}_{j})\gamma^{a}\psi_{i} + \bar{\psi}_{i}\gamma^{a}(i\partial_{\mu} \psi_{i} +{V}_{\mu}\psi_{i}-\psi_{j} W_{\mu ji})\right.\\
\left.+i\bar{\psi}_{i}\gamma^{b}\psi_{i} \Gamma_{\ b\mu}^{a}\right]\theta_{a}^{\mu}=0.
\end{eqnarray*}
%\end{widetext}
We define a Lagrangian
\begin{equation}\label{fundamentalA2}
\mathcal{L}= \bar{\psi}_{i} \gamma^{a}(i\partial_{\mu} \psi_{i}+V_{\mu}\psi_{i}-\psi_{j} W_{\mu ji})\theta^{\mu}_{a}+\frac{i}{2}\bar{\psi}_{i}\gamma^{b}\psi_{i} \Gamma_{\ b\mu}^{a}\theta_{a}^{\mu}.
\end{equation}
The last term in Lagrangian \eqref{fundamentalA2} is Yukawa coupling term $\bar{\psi}_{i}\phi \psi_{i}$ and the scalar (Higgs) field is gamma matrix valued and originatd from gravitational field
\begin{equation}\label{Higgs}
\phi=\frac{i}{2}\gamma^{b} \Gamma_{\ b\mu}^{a}\theta_{a}^{\mu}.
\end{equation}
Then, the Lagrangian \eqref{fundamentalA2} describes $U(4^{\prime})\times U(4)$ Yang-Mills theory in curved space-time.  The Lagrangian (\ref{fundamentalA2}) has relation with (\ref{fundamentalA0})
\begin{equation}
\mathbf{tr}\nabla l(x)=\mathcal{L}-\mathcal{L}^{\dagger}.
\end{equation}
If equation (\ref{fundamentalA0}) being satisfied, the Lagrangian (\ref{fundamentalA2}) is Hermitian
\begin{equation}\label{unita}
\mathcal{L}=\mathcal{L}^{\dagger}.
\end{equation}
So, the unitary principle of quantum field theory (\ref{unita}) consistents with generalized self parallel transportation principle (\ref{fundamentalA0}).
The equations of motion for the Lagrangian (\ref{fundamentalA2}) are
\begin{eqnarray}\label{fundamentalA1} 
 \gamma^{a}(i\partial_{\mu} \psi_{i}+V_{\mu}\psi_{i}-\psi_{j} W_{\mu ji})\theta^{\mu}_{a}+\frac{i}{2}\gamma^{a}\psi_{i} \Gamma_{\ a\mu}^{b} \theta_{b}^{\mu}  =0,  \ \ \ 
\end{eqnarray}
and this equation's conjugate transpose. We point out that (\ref{fundamentalA0}) is density matrix version of (\ref{fundamentalA1}). The effective equation of motion of (\ref{fundamentalA1}) has signature $(1,-1,-1,-1)$. For example, the massless Dirac equation in curved space-time of this model is
\begin{equation}\label{Dirac}
i\gamma^{a}\theta_{a}^{\mu}\partial_{\mu} \psi_{i} =0.
\end{equation}
The square of equation \eqref{Dirac} is massless Klein-Gordon equation in curved space-time
\begin{equation}\label{Klein}
\eta^{ab}\theta^{\mu}_{a}\theta^{\nu}_{b}\partial_{\mu}\partial_{\nu} \psi_{i} =0.
\end{equation}
The signature of equation \eqref{Klein} is $(1,-1,-1,-1)$ and consistents with special relativity. 

Then, a Lagrangian (\ref{fundamentalA2}) which describes the $U(4^{\prime})\times U(4)$ Yang-Mills theory in curved space-time is constructed.
Where the gravitational field $\theta^{\mu}_{a}$ is all other fields (except Higgs field) dynamical background which satisfies the characteristic of the gravitational field in our real world.

Lagrangian (\ref{fundamentalA2}) is $U(4^{\prime})\times U(4)$ gauge invariant, locally Lorentz invariant and generally covariant. So, Lagrangian (\ref{fundamentalA2}) is demanded invariant under the transformations
\begin{equation}\label{transforA}
\psi^{\prime}_{i}=\tilde{U}\psi_{j} U_{ji}, \gamma^{a\prime}=\tilde{U}\gamma^{b}\tilde{U}^{\dagger} \Lambda_{b}^{\ a}, \theta^{\prime\mu}_{a}=\Lambda^{\ b}_{a}\theta_{b}^{\mu}, 
\end{equation}
where $\tilde{U}\in U(4^{\prime})$, $(U_{ij})\in U(4)$, $\Lambda^{\ b}_{a}\in O(1,3) $, $\Lambda^{\ \mu}_{\nu}\in Gl(4,\mathbb{R}) $. Then, the transformation rules have to be derived as follows
%\begin{subequations}\label{transforB}
\begin{equation}\label{transforB3}
 V^{\prime}_{\mu}=\tilde{U}V_{\mu}\tilde{U}^{\dagger}-(\partial_{ \mu} \tilde{U}) \tilde{U}^{\dagger},
 \end{equation}
\begin{equation}\label{transforB1}
W^{\prime}_{\mu ij}=U^{*}_{ki}W_{\mu kl}U_{lj}+U^{*}_{ki}\partial_{\mu} U_{kj} ,
\end{equation}
\begin{equation}\label{transforB2}
\Gamma^{\prime b}_{\ a\mu}=\Lambda^{\ c}_{a}\Gamma^{d}_{\ c\mu}\Lambda^{\ b}_{d}-\Lambda^{\ c}_{a}\partial_{\mu}\Lambda^{\ b}_{c},
\end{equation}
%\end{subequations}
where $\tilde{U}\tilde{U}^{\dagger}=I, U^{*}_{ji}U_{jk}=\delta_{ik}$ and $\Lambda^{\ b}_{a}\Lambda^{\ c}_{b}=\delta^{c}_{a}$ are used.
Now, we complete the proof of the Lagrangian (\ref{fundamentalA2}) is $U(4^{\prime})\times U(4)$ gauge invariant, locally Lorentz invariant and generally covariant.

The gauge field strength tensors and curvature tensor are defined as follows
%\begin{subequations}
\begin{eqnarray*}\label{cur3}
% \begin{equation}
   H_{\mu\nu}&=&\partial_{\mu}V_{\nu}-\partial_{\nu}V_{\mu }-i V_{\mu}V_{\nu}+i V_{\nu}V_{\mu},\\
%\end{equation}
\label{cur2}% \begin{equation}
   F_{\mu\nu ij}&=&\partial_{\mu}W_{\nu ij}-\partial_{\nu}W_{\mu ij}-i W_{\mu ik}W_{\nu kj}+i W_{\nu ik}W_{\mu kj},\\
%\end{equation}
\label{cur}
%\begin{equation}
\label{cur1}
 R^{a}_{\ b\mu\nu}&=&\partial_{\mu}\Gamma^{a}_{\ b\nu}-\partial_{\nu}\Gamma^{a}_{\ b\mu}+\Gamma^{c}_{\ b\nu}\Gamma^{a}_{\ c\mu}-\Gamma^{c}_{\ b\mu}\Gamma^{a}_{\ c\nu},
% \end{equation}
\end{eqnarray*}
%\end{subequations}
where $R_{ab\mu\nu}=-R_{ba\mu\nu}$ if $\nabla g=0$ and $H_{\mu\nu}^{\dagger}=H_{\mu\nu}$, $F^{*}_{\mu\nu ij}=F_{\mu\nu ji}$. The gauge field strength can be decomposed by the $U(4)$ generators $H_{\mu\nu}=H_{\mu\nu}^{\alpha}\mathcal{T}^{\alpha}, F_{\mu\nu ij} = F^{\alpha}_{\mu\nu}\mathcal{T}^{\alpha}_{ij}$.
After the torsion being defined
\begin{equation}
T^{a}_{\ \nu\rho}=2\Gamma^{a}_{\ [\nu\rho]},
\end{equation}
we have the Ricci identity and Bianchi identity \cite{Fecko:2006zy} on this geometry structure as follows
%\begin{subequations}
\begin{eqnarray}
\partial_{[\mu}H_{\nu\rho] }&=&H_{[\mu\nu}V_{\rho]}-V_{[\mu}H_{\nu\rho]},\\
\partial_{[\mu}F_{\nu\rho] ij}&=&F_{[\mu\nu|ik|}W_{\rho]kj}-W_{[\mu |ik|}F_{\nu\rho] kj}, \\
          T^{a}_{\ \sigma[\rho}T^{\sigma}_{\ \mu\nu]}&=& R^{a}_{\ [\rho\mu\nu]} +\nabla_{[\rho}T^{a}_{\ \mu\nu]},\\
          \nabla_{[\rho} R^{a}_{\ |b|\mu\nu]}&=&R^{a}_{\ b\sigma[\rho}T^{\sigma}_{\ \mu\nu]}.
                    \end{eqnarray}                          
%\end{subequations}
There is Yang-Mills Lagrangian for gauge bosons in this model
\begin{equation}
\mathcal{L}_{Y}=\frac{-1}{2}\mathbf{tr}\left( H^{\mu\nu}H_{\mu\nu}\right)-\frac {\zeta}{2}F^{\mu\nu}_{ij}F_{\mu\nu ji},
\end{equation}
where $\zeta\in \mathbb{R}$ is constant.

For the gravity, the Einstein-Hilbert action in Lorentz manifold be showed as follow
\begin{equation}\label{Ein}
  S=\int \omega R  ,
\end{equation}
where $R$ is  the Ricci scalar curvature in Lorentz manifold, $\omega=\sqrt{-g_{v}} dx^{0}\wedge dx^{1}\wedge dx^{2}\wedge dx^{3}$ is volume form. And in this geometry framework, the equations can be derived as follows
%\begin{widetext}
\begin{eqnarray}  \label{relationA}
 \nabla_{[\mu}\nabla_{\nu]}l  &=&\frac{-1}{2}\left(\bar{\psi}_{i} \gamma^{a}\psi_{k}F_{\mu\nu kj}  -F^{*}_{\mu\nu ki}\bar{\psi}_{k}  \gamma^{a} \psi_{j} \right. \\    \nonumber
    &+& \bar{\psi}_{i}\tilde{H}_{\mu\nu}\gamma^{a}\psi_{j} -\bar{\psi}_{i}\gamma^{a}H_{\mu\nu}\psi_{j} +  \frac{i}{2}\left.\bar{\psi}_{i} \gamma^{b}\psi_{j} R^{a}_{\ b\mu\nu}\right)e^{\dagger}_{j} \otimes e_{i}\theta_{a},\\    \nabla_{[\mu}\nabla_{\nu]}\tilde{l}  &=&\frac{-1}{2}\left(\bar{\psi}_{i} \gamma^{a\dagger}\psi_{k}F_{\mu\nu kj}  - F^{*}_{\mu\nu ki}\bar{\psi}_{k}  \gamma^{a\dagger} \psi_{j} \right. \\   \nonumber
   &+& \bar{\psi}_{i}H_{\mu\nu}\gamma^{a\dagger}\psi_{j} -\bar{\psi}_{i}\gamma^{a\dagger}\tilde{H}_{\mu\nu}\psi_{j} +  \frac{i}{2}\left.\bar{\psi}_{i} \gamma^{b\dagger}\psi_{j} R^{a}_{\ b\mu\nu}\right)e^{\dagger}_{j} \otimes e_{i}\theta_{a},
\end{eqnarray}
where $\tilde{H}_{\mu\nu}=\gamma^{0}H_{\mu\nu}\gamma^{0}$.
%\end{widetext}
We define $ \nabla^{2}=\nabla_{[\mu}\nabla_{\nu]} dx^{\mu}\wedge dx^{\nu}$, the equation of this gravity theory is constructed
\begin{equation}\label{gravity}
  \mathbf{tr} \nabla^{2} [\tilde{l}(x) l(x)]=0.
\end{equation}
This equation (\ref{gravity}) is obviously $U(4^{\prime})\times U(4)$ gauge invariant, locally Lorentz invariant and generally covariant. The explicit formula of equation (\ref{gravity}) is
\begin{equation}\label{gravity0}
R=\frac{i}{4}\left(F_{abij}\psi^{\dagger}_{j}(\gamma^{a}\gamma^{b}-\gamma^{b\dagger}\gamma^{a\dagger})\psi_{i}-H_{ab}(\gamma^{a}\gamma^{b}-\gamma^{b\dagger}\gamma^{a\dagger})\right),
\end{equation}
where $\partial_{\mu}dx^{\nu}\otimes dx^{\rho} \partial_{\sigma}=\delta_{\mu}^{\nu} \delta_{\sigma}^{\rho}$, $dx^{\mu}\otimes dx^{\nu} \partial_{\rho}\partial_{\sigma}=\delta^{\nu}_{\rho} \delta_{\sigma}^{\mu}$ are used and $F_{abij}=F_{\mu\nu ij} \theta_{a}^{\mu}\theta_{b}^{\nu}, H_{ab}=H_{\mu\nu} \theta_{a}^{\mu}\theta_{b}^{\nu}$. So we define a $U(4^{\prime})\times U(4)$ gauge invariant, locally Lorentz invariant, generally covariant Lagrangian
\begin{eqnarray}  \label{Lg}\nonumber
\mathcal{L}_{g}=R \psi^{\dagger}_{i} \psi_{i}-i\left(F_{abij}\psi^{\dagger}_{j}(\gamma^{a}\gamma^{b}-\gamma^{b\dagger}\gamma^{a\dagger})\psi_{i}-\psi^{\dagger}_{i}H_{ab}(\gamma^{a}\gamma^{b}-\gamma^{b\dagger}\gamma^{a\dagger})\psi_{i}\right). \\ 
\end{eqnarray}
The Lagrangian (\ref{Lg}) is Hermitian
\begin{equation}
\mathcal{L}_{g}=\mathcal{L}^{\dagger}_{g}.
\end{equation}
The $R\psi^{\dagger}_{i} \psi_{i}$ in Lagrangian (\ref{Lg}) gives us the Einstein-Hilbert action. The equation (\ref{gravity0}) and the Einstein tensor can be derived from the Einstein-Hilbert action. 
\section{Sheaf quantization and path integral quantization}
The entities $l(x)$ and $\tilde{l}(x)$ are two sections of the two bundles, respectively, where these two bundles are dual to each other. Further, the Sheaf valued entities $\hat{l}(x)$ and $\hat{\tilde{l}}(x)$ can be defined
\begin{eqnarray}\label{lhat}
\hat{l}(x)=\sum_{\kappa}\eta_{\kappa}(x)|\kappa,x\rangle \langle \kappa,x| l_{\kappa}(x),\\
\hat{\tilde{l}}(x)=\sum_{\kappa}\eta_{\kappa}(x)|\kappa,x\rangle \langle \kappa,x| \tilde{l}_{\kappa}(x),
\end{eqnarray}
 where $\eta_{\kappa}(x)\in [0,1]$ are probability of corresponding section $l_{\kappa}(x)$ and $\tilde{l}_{\kappa}(x)$, $\kappa$ is Sheaf space index and evaluated in an abelian group.
 The density matrix corresponds to $\hat{l}(x)$ and $\hat{\tilde{l}}(x)$ is
\begin{equation}\label{lhat}
\rho(x)=\sum_{\kappa}\eta_{\kappa}(x)|\kappa,x\rangle \langle \kappa,x|.
\end{equation}
 We have orthgonal bases in Sheaf space and probability complete formulas 
 \begin{eqnarray}  \label{oth}
\langle \kappa,x|\kappa^{\prime},x^{\prime}\rangle&=&\delta(x-x^{\prime})\delta(\kappa-\kappa^{\prime}) ,\\ 
\label{pro_c}
\mathbf{tr}\rho(x)&=&\sum_{\kappa}\eta_{\kappa}(x)=1.
\end{eqnarray}
In mathematic, a Sheaf is a collection of sections, the index $\kappa$ of each section correspond to an abelian group element. In physics, the Sheaf spaces $Sh(x)$ and $\tilde{Sh}(x)$ are expanded by all possible sections $l_{\kappa}(x)$ and $\tilde{l}_{\kappa}(x)$ of the two bundles, respectively. The $Sh(x)$ and $\tilde{Sh}(x)$ are linear spaces, which means, for example, any two entitis in $Sh(x)$, there is a entity equals to the mixing of the two entitis
\begin{eqnarray}\nonumber
&&\hat{l}(x)=\eta_{1}(x) \hat{l}_{1}(x)+\eta_{2}(x)\hat{l}_{2}(x) ;   \quad  \hat{l}_{1}(x), \hat{l}_{2}(x) \in Sh(x)\\
&\Rightarrow& \hat{l}(x)\in Sh(x),
\end{eqnarray}
 where $\eta_{1}(x), \eta_{2}(x)\in [0,1]$ and $\eta_{1}(x)+\eta_{2}(x)=1$.
 The Sheaf spaces $Sh(x)$ and $\tilde{Sh}(x)$ are dual to each other.
We call it Sheaf quantization which switching study objects from single section to all possible sections of the bundle.
The equation (\ref{pro_c}) derives to equation of motion for density matrix
\begin{equation}\label{lhat}
d (\mathbf{tr}\rho(x))=\mathbf{tr} (d\rho(x))=0,
\end{equation}
where $d$ is exterior differential derivative.
The equations of motion for entities $\hat{l}(x)$ and $\hat{\tilde{l}}(x)$ after Sheaf quantization are
 \begin{equation}\label{fundamentalA00}
\mathbf{tr}\nabla [\hat{l}(x)]= 0,  \quad
 \mathbf{tr} \nabla^{2} [\hat{\tilde{l}}(x) \hat{l}(x)]=0.
\end{equation}
The corresponding total Lagrangian density is
\begin{eqnarray}
\hat{\mathcal{L}}=\sum_{\kappa}\eta_{\kappa}(\mathcal{L}_{\kappa}+g\mathcal{L}_{g,\kappa}+\tilde{g}\mathcal{L}_{Y,\kappa}),
\end{eqnarray}
where $g,\tilde{g}$ are Lagrange multipliers and $g,\tilde{g}\in\mathbb{R}$. 

For pure state
\begin{eqnarray}
\rho(x)&=&|\Psi(x)\rangle\langle \Psi(x)|,\\
|\Psi(x)\rangle&=&\sum_{\kappa}\alpha_{\kappa}(x) |\kappa\rangle,
\end{eqnarray}
where the $|\Psi(x)\rangle$ is the quantum state of quantum field theory.
The transition amplitude can be defined through path integral formula
\begin{eqnarray}
\alpha_{\kappa}(t,\vec{x})=\int_{t^{\prime}\in (t_{0},t)} D\kappa(t^{\prime},\vec{x}) e^{i\int \omega \hat{\mathcal{L}}[\kappa(t^{\prime},\vec{x})]} \alpha_{\kappa}(t_{0},\vec{x}). \ \
\end{eqnarray}

\begin{figure}
\begin{center}
  \includegraphics[width=86mm]{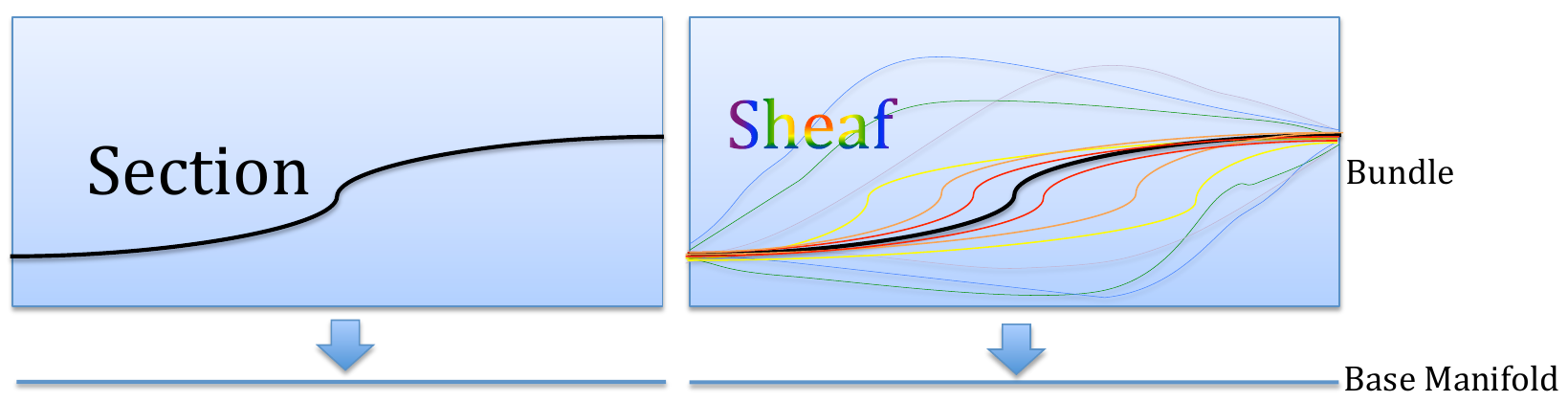}
  \caption{\label{figure4} Left: The fiber bundle structure and $l(x)$ is a section of the bundle.  Right: A Sheaf is a collection of the sections. $\hat{l}(x)$ is Sheaf valued.}
\end{center}
\end{figure}  

%For historical reasons, this process is called second quantization because this process switch us from single particle free motion to multi-particles interaction in sometimes. 

\section{Particles Spectrum}
\label{sec:2}
$V_{\mu}^{\alpha}$ and $W_{\mu}^{\alpha}$ ($\alpha=0,1,\cdots, 15$) are gauge bosons fields.  The interactions related with $W_{\mu}^{\alpha}$ always preserves the possibility of chiral symmetry breaking such that the gauge group can decomposed to $U(4^{\prime})\times U(4)_{L}\times U(4)_{R}$, where $U(4^{\prime})$ is color group and $U(4)_{L}\times U(4)_{R}$ is chiral flavor group.
The $V_{\mu}^{0}$ is dark photon and $W_{\mu}^{0}$ is Fiona particle. The left over part gauge group is a Pati-Salam gauge group $SU(4^{\prime})\times SU(4)_{L}\times SU(4)_{R}$ \cite{Pati:1974yy,Li2020zij} and the $SU(4^{\prime})$ can be decomopsed as follow
 \begin{equation}\label{SU4}
  SU(4^{\prime})=SU(3^{\prime})\oplus  U(1^{\prime})+ U_{X^{+}}+ U_{X^{-}}.
 \end{equation}
 The $SU(3^{\prime})$ is the gauge group of quantum chramodynamics (QCD) and the corresponding gauge bosons $V_{\mu}^{\alpha}(\alpha=1,2\cdots,8)$ are gluons. The $U(1^{\prime})$ is electro-magnetic interaction gauge group and corresponding gauge boson $V_{\mu}^{15}$ is photon $\gamma$. The $X^{\pm}$ particles transport semi-leptonic processes and 
  \begin{equation}\label{X}
  X^{\pm C}=V_{\mu}^{8+C}\pm iV_{\mu}^{9+C}.
 \end{equation}
 The electric charge of $X^{+}$ and $X^{-}$ are $\frac{1}{3}$ and $-\frac{1}{3}$.
 The chiral gauge group $SU(4)_{L,R}$ can be decomposed as 
  \begin{equation}\label{SU4_Chiral}
  SU(4)_{L,R}=SU(3)_{Y}\oplus  U(1)_{Z}+ U_{W^{+}}+ U_{W^{-}},
 \end{equation}
 and related gauge bosons $W_{\mu}^{\alpha}(\alpha=1,2,\cdots,15)$ contain weak bosons $W^{\pm}$ and $Z$
 \begin{eqnarray}  \label{weakzp}
W^{\pm}_{\mu}&=&W^{9}_{\mu} \pm i W^{10}_{\mu}=W^{11}_{\mu} \pm i W^{12}_{\mu}=W^{13}_{\mu} \pm i W^{14}_{\mu},\ \  \\   \label{weakzp}
\label{weakw}
Z_{\mu}&=&W^{15}_{\mu}.
\end{eqnarray}
The left over gauge bosons are $Y^{0},Y^{1},Y^{2}$ and $Y^{1}_{*},Y^{2}_{*}$ with 0 eletric charge. The gauge bosons $Y^{1},Y^{2},Y^{1}_{*},Y^{2}_{*}$ transport non-SM flavor changing neutral currents (FCNCs) and
\begin{eqnarray}
&&\left(\begin{array}{ccc}
(2-\eta)Y^{0}_{\mu}& Y_{\mu}^{1}&Y_{\mu}^{2}\\
Y_{*\mu}^{1}& \eta Y^{0}_{\mu}&Y_{\mu}^{1}\\
Y_{*\mu}^{2}& Y_{*\mu}^{1}&-2 Y^{0}_{\mu}
\end{array}\right)=2\sum_{\alpha=1}^{8}W_{\mu}^{\alpha}\mathcal{T}^{\alpha}.
\end{eqnarray}
The $X^{\pm}$ and $Y^{1},Y^{2},Y^{1}_{*},Y^{2}_{*}$ must be superheavy from the restrictions of experimental data.
\begin{figure}
\begin{center}
  \includegraphics[width=90mm]{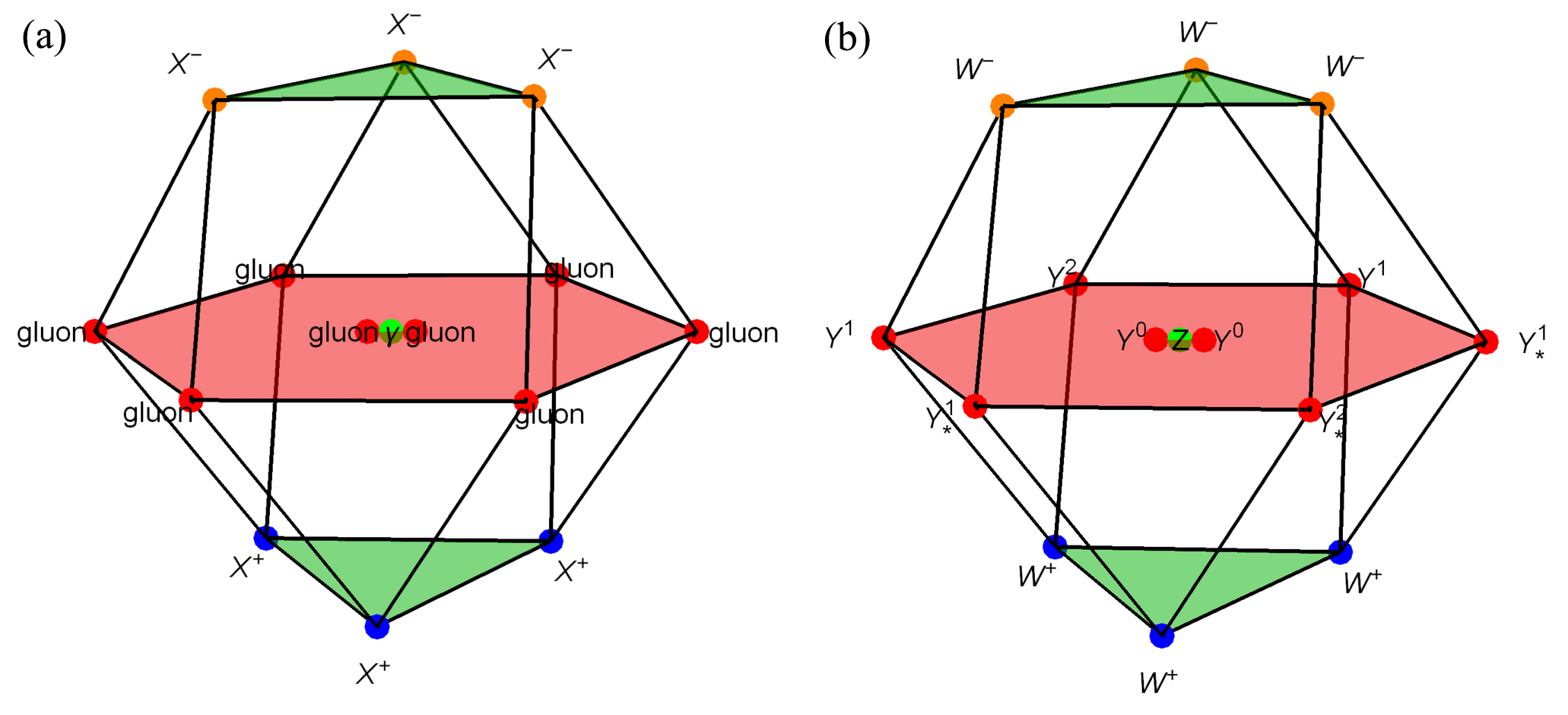}
  \caption{\label{figure1}Weight diagram of $SU(4)$ adjoint representation and corresponding gauge bosons. The decomposition of $SU(4)$ adjoint representation is $\mathbf{15}=\mathbf{8}\oplus \mathbf{1}+\mathbf{3}+\mathbf{3^*}$. (a) The wight diagram of $V_{\mu}^{\alpha}(\alpha=1,2,\cdots, 15)$ related gauge bosons. (b) The wight diagram of $W_{\mu}^{\alpha}(\alpha=1,2,\cdots, 15)$ related gauge bosons.}
\end{center}
\end{figure}

\begin{figure}
\begin{center}
  \includegraphics[width=78mm]{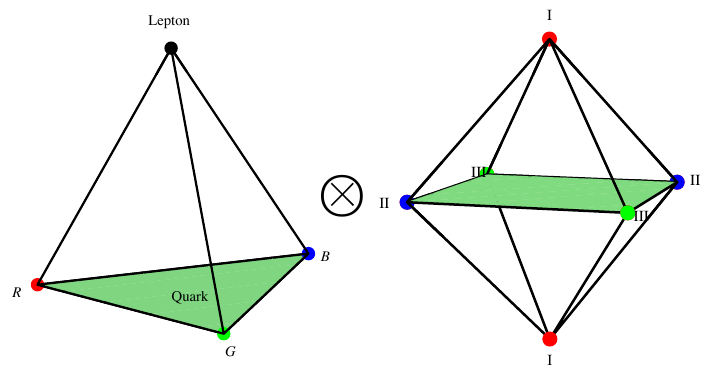}\\
  \caption{\label{figure2}Weight diagram of $SU(4)$ fundamental representation $\mathbf{4}\otimes \mathbf{6}$. Representation $\mathbf{4}=\mathbf{3}+\mathbf{1}$, $\mathbf{3}$ is 3 kinds of color, red, green and blue, $\mathbf{1}$ is the lepton number. Representation $\mathbf{6}$ give us 6 flavor of quarks and leptons. 6 flavor has 3 generations, \uppercase\expandafter{\romannumeral1} ,\uppercase\expandafter{\romannumeral2} and \uppercase\expandafter{\romannumeral3}, each generation has 2 kinds of quarks or leptons.}
  \end{center}
\end{figure}

The fermionic fields $\psi_{i}$ transform as the $U(4^{\prime})\times U(4)$ fundamental representation according to (\ref{transforA}).
So, fermions are filled into the $SU(4)$ fundamental representation $\mathbf{4}\otimes \mathbf{6}$ naturally. The fundamental representation $\mathbf{4}$ corresponds to 3 colors and 1 lepton and leads us reobtain ``Lepton number as the fourth color'' \cite{Pati:1974yy} . The fundamental representation $\mathbf{6}$ corresponds to 6 flavor of quarks and leptons.   The weight diagram coordinates in Chevalley basis of representation $\mathbf{6}$ (see in TABLE.~\ref{table1})  have good correspondence with the quarks quantum number \cite{Agashe:2014kda}. The antifermions be filled into the representation $\mathbf{6}\otimes\mathbf{\bar{4}}$ similarly.   The weight diagram of fermions is shown in FIG.~\ref{figure2}.
Both left handed and right handed fermions for all quarks and leptons are existed. Especially, the existence of right handed neutrinos is predicted. This is compatible with experimental results \cite{Fukuda:1998fd} and the well know Seesaw mechanism \cite{Minkowski:1977sc,Mohapatra:1979ia,Yanagida:1980xy,Schechter:1980gr,Xing:2011zza}. The $2I_{z}$ being used in Table~\ref{table1} are not no reasons because we have  the Gell-Mann-Nishijima formula \cite{Agashe:2014kda}
\begin{equation}
  Q=I_{z}+\frac{\mathcal{B}+S+C+B+T}{2}.
\end{equation}

\begin{table}
\caption{\label{table1}Corresponding relations between quarks quantum number and weight diagram coordinates in Chevalley basis $H_{1},H_{2}$ and $H_{3}$ of $SU(4)$ representation $\mathbf{6}$.}
\begin{ruledtabular}
\begin{center}
\begin{tabular}{ccccccc}    %\hline \hline
&2$I_{z}$ & S+C & B+T & $H_{1}$ & $H_{2}$ & $H_{3}$\\ \hline
 u&1 & 0 & 0 & 1 & -1 & 1\\
 d&-1 & 0 & 0 & -1 & 1 & -1\\
 c&0 & 1 & 0 &1 & 0 & -1\\
 s&0 & -1 & 0 &-1 & 0 & 1 \\
 t&0 & 0 & 1 & 0 & 1 & 0\\
 b&0 & 0 & -1 &0 & -1 & 0\\  %\hline \hline
\end{tabular}
\end{center}
\end{ruledtabular}
\end{table}

An explicit fermions representation \cite{Li2020zij} in this model might be
\begin{eqnarray}\label{representation}
\psi_{i}=\left(\begin{array}{cccc}
\sqrt{2}u_{R}& \sqrt{2}c_{R} & \sqrt{2}t_{R} &d^{\prime}_{R}\\
\sqrt{2}u_{G}& \sqrt{2}c_{G} & \sqrt{2}t_{G} &d^{\prime}_{G}\\
\sqrt{2} u_{B}& \sqrt{2}c_{B} & \sqrt{2}t_{B} &d^{\prime}_{B}\\
e & \mu &\tau & \nu^{\prime}
\end{array}\right),
\end{eqnarray}
where $u,c,t$ and $d^{\prime}$ are quarks fields, $e,\mu,\tau$ and $\nu^{\prime}$ are electron, mu, tau and neutrinos fields.  The corresponding fermions electric charges of (\ref{representation}) are
\begin{eqnarray}
Q=\left(\begin{array}{cccc}
2/3& 2/3 & 2/3 & -1/3\\
2/3& 2/3 & 2/3 & -1/3\\
2/3& 2/3 & 2/3 & -1/3\\
1 & 1 & 1 & 0
\end{array}\right).
\end{eqnarray}
 The quarks states like $|d\rangle, |s\rangle,|b\rangle$ and neutrinos states $|\nu_{e}\rangle, |\nu_{\mu}\rangle,|\nu_{\tau}\rangle$ are eigen states of the Lagrangian \cite{Li2020zij}.

\section{Conclusion and Discussion}
\label{sec:3}
A Pati-Salam model and a gravity theory from square root Lorentz manifold are derived. A Sheaf quantization scheme which consistents with path integral quantization is shown. The particles spectrum in this model is dicussed.

Some possible new physics on this model are listed as follows. There are exotic gauge bosons such as dark photon, Fiona, $X^{\pm}$ and $Y^{0},Y^{1},Y^{2}, Y^{1}_{*},Y^{2}_{*}$. The $X^{\pm}$ transports semi-leptonic processes, the $Y^{1},Y^{2}, Y^{1}_{*},Y^{2}_{*}$ transport non-SM FCNCs. The right handed neutrinos are existed. The Higgs field is gamma matrix valued.

\begin{acknowledgments}
We thank Professor Chao-Guang Huang for long time discussions and powerful help. Without Professor Huang's help, this work is impossible. We thank Zhao Li, Cai-Dian Lv, Jun-Bao Wu, Ming Zhong, Yong-Chang Huang, Qi-Shu Yan, Yu Tian, De-Shan Yang, Yang-Hao Ou for valuable discussions. We thank Jean Thierry-Mieg for critical comments. We thank Yu Lu for computer drawing.
\end{acknowledgments}

\appendix

\section{Generators of $U(4)$}

$\mathcal{T}^1=\frac{1}{2}\begin{pmatrix}
   0&1&0&0\\
   1&0&0&0\\
   0&0&0&0\\
   0&0&0&0\\
    \end{pmatrix},
 \mathcal{T}^2=\frac{1}{2}\begin{pmatrix}
   0&-i&0&0\\
   i&0&0&0\\
   0&0&0&0\\
   0&0&0&0\\
    \end{pmatrix}, \\
    \mathcal{T}^3=\frac{1}{2}\begin{pmatrix}
   1&0&0&0\\
   0&-1&0&0\\
   0&0&0&0\\
   0&0&0&0\\
    \end{pmatrix},
 \mathcal{T}^4=\frac{1}{2}\begin{pmatrix}
   0&0&1&0\\
   0&0&0&0\\
   1&0&0&0\\
   0&0&0&0\\
    \end{pmatrix},\\
\mathcal{T}^5=\frac{1}{2}\begin{pmatrix}
   0&0&-i&0\\
   0&0&0&0\\
   i&0&0&0\\
   0&0&0&0\\
    \end{pmatrix},
 \mathcal{T}^6=\frac{1}{2}\begin{pmatrix}
   0&0&0&0\\
   0&0&1&0\\
   0&1&0&0\\
   0&0&0&0\\
    \end{pmatrix}, \\
\mathcal{T}^7=\frac{1}{2}\begin{pmatrix}
   0&0&0&0\\
   0&0&-i&0\\
   0&i&0&0\\
   0&0&0&0\\
    \end{pmatrix},
 \mathcal{T}^8=\frac{\sqrt{3}}{6}\begin{pmatrix}
   1&0&0&0\\
   0&1&0&0\\
   0&0&-2&0\\
   0&0&0&0\\
    \end{pmatrix},\\
\mathcal{T}^9=\frac{1}{2}\begin{pmatrix}
   0&0&0&1\\
   0&0&0&0\\
   0&0&0&0\\
   1&0&0&0\\
    \end{pmatrix},
 \mathcal{T}^{10}=\frac{1}{2}\begin{pmatrix}
   0&0&0&-i\\
   0&0&0&0\\
   0&0&0&0\\
   i&0&0&0\\
    \end{pmatrix},  \\
\mathcal{T}^{11}=\frac{1}{2}\begin{pmatrix}
   0&0&0&0\\
   0&0&0&1\\
   0&0&0&0\\
   0&1&0&0\\
    \end{pmatrix},
 \mathcal{T}^{12}=\frac{1}{2}\begin{pmatrix}
   0&0&0&0\\
   0&0&0&-i\\
   0&0&0&0\\
   0&i&0&0\\
    \end{pmatrix},\\
\mathcal{T}^{13}=\frac{1}{2}\begin{pmatrix}
   0&0&0&0\\
   0&0&0&0\\
   0&0&0&1\\
   0&0&1&0\\
    \end{pmatrix},
 \mathcal{T}^{14}=\frac{1}{2}\begin{pmatrix}
   0&0&0&0\\
   0&0&0&0\\
   0&0&0&-i\\
   0&0&i&0\\
    \end{pmatrix},  \\
 \mathcal{T}^{15}=\frac{\sqrt{6}}{12}\begin{pmatrix}
   1&0&0&0\\
   0&1&0&0\\
   0&0&1&0\\
   0&0&0&-3\\
    \end{pmatrix},
     \mathcal{T}^0=\frac{1}{2\sqrt{2}}\begin{pmatrix}
   1&0&0&0\\
   0&1&0&0\\
   0&0&1&0\\
   0&0&0&1\\
    \end{pmatrix} .$
 %   \end{widetext}
\\
\nocite{*}
%\bibliography{mybibfile.bib}% Produces the 


\end{document}